# Protected QR Code-based Anti-counterfeit System for Pharmaceutical Manufacturing

Md Masruk Aulia[a,1], Nitol Saha[b,*,1], Md. Mostafizur Rahman[c,1]

[a]*Military Institute of Science and Technology (MIST), Dhaka, 1216, Bangladesh*
[b]*Department of Mechanical Engineering, University of South Carolina, Columbia, 29201, South Carolina, USA*
[c]*Khulna University of Engineering & Technology, Khulna, 9203, Bangladesh*

* Corresponding author. Tel.: +1-803-269-8057; *E-mail address:* nsaha@email.sc.edu

**Abstract**

The pharmaceutical manufacturing faces critical challenges due to the global threat of counterfeit drugs. This paper proposes a new approach of protected QR codes to secure unique product information for safeguarding the pharmaceutical supply chain. The proposed solution integrates secure QR code generation and encrypted data transmission to establish a comprehensive anti-counterfeit ecosystem. The protected QR codes encapsulate product information that cannot be identified using traditional QR code scanners which protect the information against replication and tampering. The system is developed with scalability in mind, which can be easily implemented without introducing any additional modification in the traditional supply chain.

## 1. Introduction

Pharmaceutical counterfeiting is becoming a serious problem worldwide. Counterfeit drugs pose a serious impact on public health. Since the 1990s, there has been evidence of pharmaceutical counterfeiting, and lately, the issue has become more serious. Not just in developing nations but also, and increasingly so, in developed nations, there are a great deal more cases. Counterfeit drugs are those that are "deliberately mislabelled with respect to identity and/or source," according to the World Health Organization (WHO) [1], [2]. Both branded and generic goods are subject to counterfeiting; examples of such goods include medications that have the right or incorrect ingredients, no active ingredients, too few active ingredients, or phony packaging [3].

Counterfeiting of drugs and pharmaceutical products creates a negative impact on the economy, and health as well as seriously affects the pharmaceutical manufacturing sector. To prevent counterfeiting, manufacturers are adopting anti-counterfeiting measures to prevent counterfeiting. Holograms, barcodes, etc. have been lately used to prevent counterfeiting but failed to eradicate it completely as counterfeiters are adopting different techniques to easily tamper holograms and barcodes. The technologies used to combat counterfeiting are covered in [4] including track and trace systems, 2D matrices, QR codes, and RFID. While 2D barcodes are becoming more popular in Europe, developed nations like the USA have adopted RFID technology. Governments of different countries have created regulations requiring barcodes as they are becoming more aware of the severity of the issue. Upconversion (UC) fluorescent three-dimensional (3D) quick response (QR) code method supported by smartphone recognition has been implemented for tracking and anti-counterfeiting of drugs. A printer has been used to print a series of colors by precisely regulating the overlap of these three inks on the product i.e. Capsules [5]. Each of the three unique color layers that comprised the 3D QR code provided details about a different aspect of the drug. A smartphone application has been developed to extract product-related information from the QR code. Product packaging utilizing serialized QR codes has drawn a lot of attention as a possible remedy for the issue of industrial counterfeiting [6]. The problem related to QR codes is that they can be easily cloned to counterfeit. The technique employed in [6] used a copy-detection pattern technique which has been available in the market for several years. Copy detection pattern technique depends on information loss during replication of the printed QR code. However, this technique faces challenges of printing and scanning variability. It is challenging to implement it in small products and smartphone cameras often fail to capture the correct information due to the low-quality image. During bulk production, printing variability is common, and calibrating the printer during continuous production is not feasible as it will create huge production losses. A potential answer to this issue is offered by radio frequency identification (RFID) technology, which equips every product with a secure tag that is hard to counterfeit. Authors of [7] discussed how manufacturing companies can use Radio Frequency Identification (RFID) as a powerful technology to improve asset management, inventory control, and supply chain management. This study [7] presents an Intelligent RFID-based Electronic Anti-Counterfeit System (InRECS) that incorporates a Case-based Reasoning (CBR) engine and data mining technique to address the counterfeit issue. InRECS provides intelligent feedback into inventory and materials transfer operations along with accurate worldwide supply chain visibility. Physically Unclonable Functions (PUF) and RFID are mentioned in [8] as a means of avoiding counterfeiting. The suggested remedy is predicated on merging PUF technology with the Rabin public-key encryption technique. However, there are many security risks associated with RFID technology. For instance, if the communication link between the reader and the tag is breached, a malicious opponent may be able to access the sensitive information kept on the device [7]. These implementations also pose interoperability challenges. Careful consideration and standardization of these implementations are required which are quite challenging. In [9], the authors proposed a Blockchain technique to prevent counterfeiting. However, this technology is not yet realistic, and it requires major modification in the supply chain infrastructure which is not feasible. To prevent counterfeiting, the U.S. Drug Supply Chain Security Act (DSCSA) enacted in 2013 mandates a comprehensive track-and-trace solution for pharmaceuticals to enhance supply chain security [10]. This solution involves serialization at the individual saleable drug unit level, ensuring that each product is uniquely identified and traceable throughout the supply chain. Trading partners are required to exchange transaction information and transaction statement(s) electronically in a secure and interoperable manner, enabling efficient verification of product identifiers and facilitating unit-level traceability, thus safeguarding the pharmaceutical supply chain from counterfeit or illegitimate products.

[1] Authors with equal contribution

In [11], a QR code created can conceal information from a conventional scanner, even if it hasn't been used in the detection system of counterfeit goods yet. To stop a conventional scanner from scanning more characters, the library utilizes a null character separator. However, if a scanner is forced to scan deeper, the approach does not provide appropriate security for the data adjacent to the null character.

Based on the above literature review and taking into consideration the U.S. Drug Supply Chain Security Act (DSCSA), the proposed system described in this paper presents an innovative solution to solve the counterfeiting issues. A protected QR code-based method has been proposed where information of the serialized QR code cannot be identified using the regular QR code scanner which prevents the information leak. Every product on the market will be given a distinctive tag in the form of a QR code. The QR codes are different from the traditional QR codes. The tag will be created and encoded into the QR code images in a way that prevents its value from being retrieved by a conventional scanner. To check the product's origin and other metadata, the end user must scan the product using a special mobile application with a remote authentication server that can read the modified QR code where the location of the scanning is tracked to identify the potential location of the counterfeiting.

This system does not work traditionally like the other anticounterfeit systems which rely on singular functions like only barcode, RFID, blockchain, holograms, etc. Holograms are not computerized solutions, and they can easily mislead as they cannot be viewed from all viewing angles. On the other hand, QR code scanning is a sustainable approach that is more accurate and faster. The QR codes can store more information than barcodes. Having a segment of error correction, QR codes are reliable solutions to recover information. Compared to the blockchain method, it is simpler and more user-friendly. In metallic and conductive objects, RFID scanning can be a problem whereas it can be overcome by QR codes as optical scanning is used. The tag of QR code generation is based on the product requirements. There are several subsections: one section is responsible for tag printing, checking, and product sorting and another section is accountable for maintaining the communication between the user and the server. These two sections are interconnected with each other. Consequently, these two sections indicate the robustness of the proposed system, and the other interesting factor is the data decoding system. The secureness of data by the proposed system is a unique approach which opens innovative ideas of research about the system. The general user cannot access all the data without authenticated software. Hence, the data cannot be trespassed easily and cannot be used for reverse engineering of the system. Based on the highlighted factors like initial checking, interconnected servers, data security, and less risk of counterfeiting, it can be stated that the proposed system has the potential to overcome the critical issues and main challenges faced in the field of counterfeiting, and ultimately contributing to the prevention of counterfeiting.

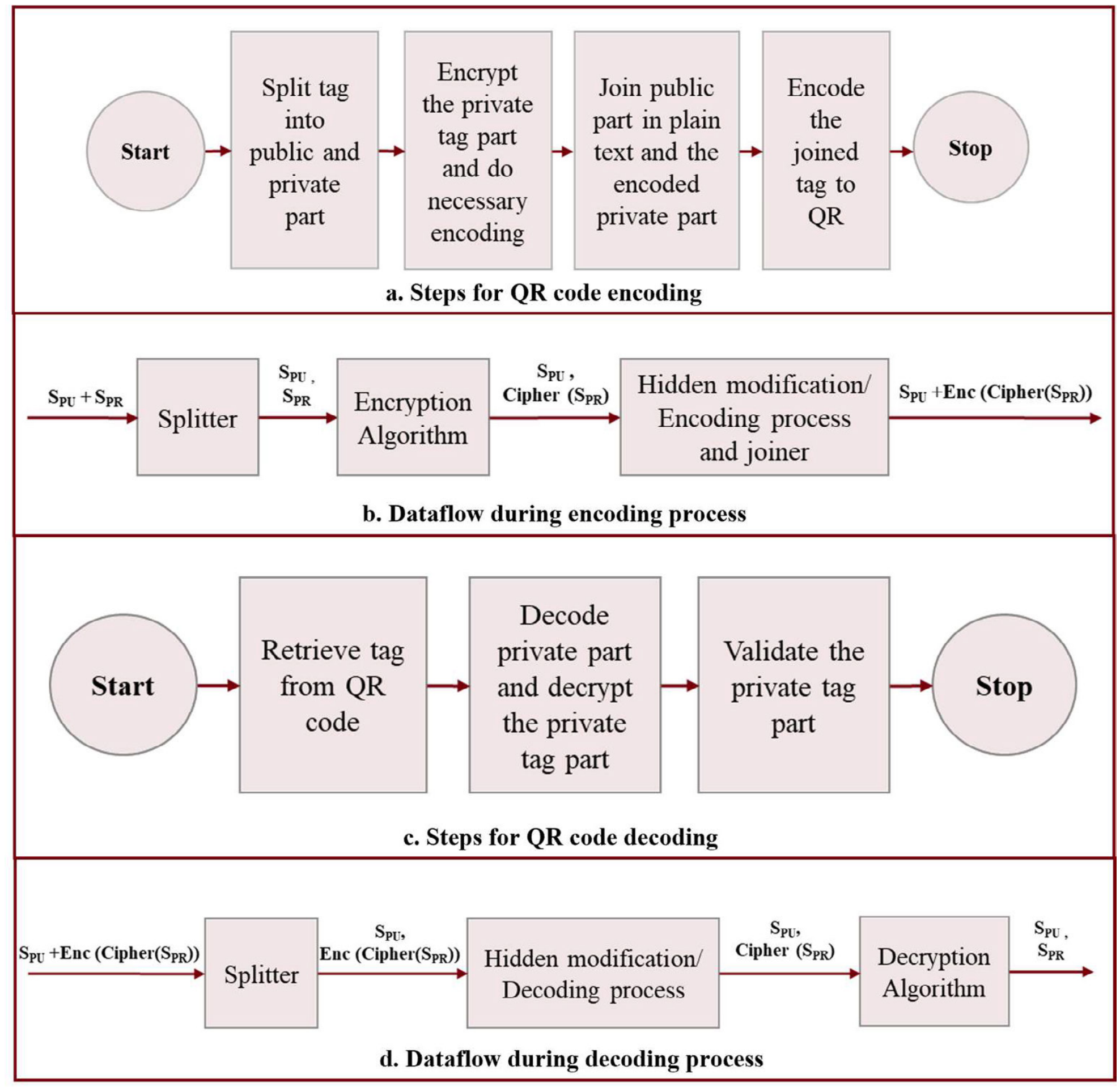

Fig. 1. Protected QR code encoding and decoding steps

## 2. System architecture

### *2.1. Protected QR code structure*

Fig. 1. depicts the process of generating the customized QR code. The method used in [11] was modified to develop the protected QR code. To modify the existing QR code, the tag string is divided into two strings: public and private string before encoding it into the QR code:

- **Public string ($S_{PU}$):** The first part is the public string which is visible to every conventional QR code scanner. This part may be empty or have some ASCII characters that partially give some metadata about the product that is not sensitive. The metadata will contain general information about the product, such as product name, batch number, manufacturer name, manufacturing date, and expiry date. This information is the same for all products of the same type.

- **Private string ($S_{PR}$):** The other and most significant part of this QR code is the private string, which contains a unique code. The unique code is converted to non-ASCII form through a hidden modification and encrypted shown in Fig. 1. (b), so this part is not readable in a conventional QR code scanner. A special program with decoding logic can only read the private part. So, the private string is the prime indicator of the product's origin. Fig. 1. (a) shows the process of the encryption of the private string.

### *2.2. Protected QR code generation lifecycle*

The protected code generation lifecycle is depicted in Fig. 2. The implementation goes through the following steps:

- **Tag generation and encoding:** Before encoding, the tags are split into private and public parts. Firstly, the private tag is encrypted using any popular encryption algorithm like RSA and DSA shown in Fig. 1. (a) [12]. The data flow through the encoding process is depicted in Fig. 1. (b). After encryption, the cipher text is modified with the help of private code in a way so that it is no longer readable

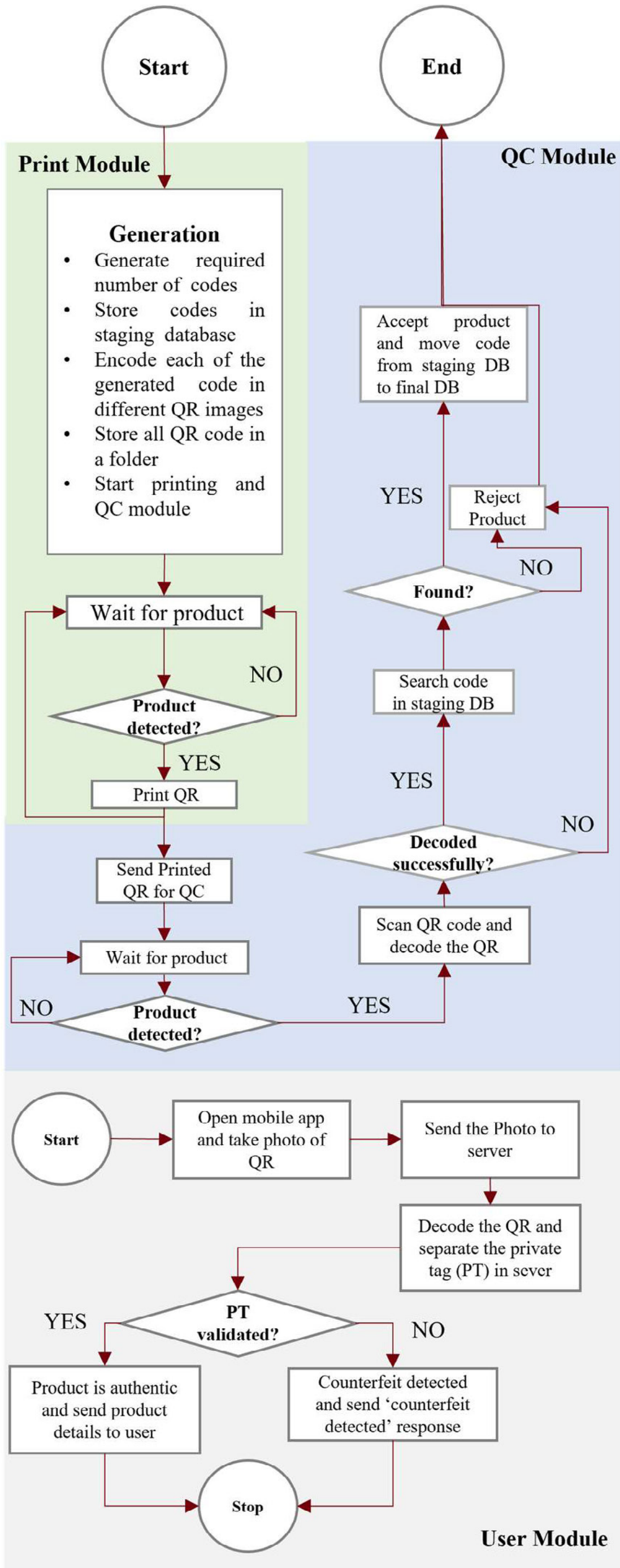

Fig. 2. Protected QR code generation lifecycle

ASCII string. This process of modification is secret which adds a layer of security to the private tag. Finally, the non-readable string is merged with the readable public part and encoded into the QR image. Then the tags are saved in a database. Each tag is encoded to different QR images. This way the QR code is modified by adding a different dimension to the existing QR code methodology. Every individual product from a company will be assigned a unique tag. The required number of QR images is generated based on the product quantity. All the images are stored in a folder, and image details like their path, tag, and other product-related information are stored in a database. This is called a staging database, where data is stored temporarily.

- **Decoding:** A special decoding program is integrated with the tag generator software or can be deployed on a local server for faster processing of several products during printing. An integrated program takes a shorter processing time for the image than a deployed API service. However, the deployed API increases the security of the decoding algorithm. Both of the methods can be implemented for robustness. The decoder program decodes the tag from the QR and splits the tag into private and public parts. The tag will be scanned two times: the first time during the quality control test after printing the code on a product in the factory. Then, during the verification of products at the user end. The decryption and decoding process of the private string is depicted in Fig. 1. (c) and the dataflow during the decoding is depicted in Fig. 1. (d). The encrypted non-readable string is retrieved from the QR. Following the encrypted data's recovery, it is initially decoded using the concealed modification function before being decrypted. The tag is then finally readable by decoding with the help of previous cipher text.

- **Quality control:** After the generation, the images are ready to be printed. As soon as a sensor detects a product's arrival, it fetches one image based on the ordered data in the database and prints the image on the product. The printed image might be distorted and not clear enough to retrieve data from it. So, all the products go through a QC test before they go to market. A camera captures a photo of the printed product when another sensor detects it. Then the image is decoded locally or from a server. If the image cannot be decoded or the decoded data is not found in the staging database, the product is rejected. The product is accepted if and only if decoded data is found in the staging database. When a product is accepted, all of its information is moved to the final database. The client application will use the final database for further verification and data retrieval.

- **QR code authentication:** The end user needs a special mobile application to scan the QR code printed on the product. The decoder is not integrated with the application to decode the QR code locally as it has security threats. If the decoder is integrated with the application, the decoding steps and keywords have the risk of being revealed by reverse engineering, and sending decoded data over the internet is not completely safe. To avoid these threats, the application captures the image only with the location where it was scanned and sends it to the server. The server will decode the image, verify the image, and send a response message to the application.

## 3. Implementation

The implementation architecture of the proposed system is shown in Fig. 3. The components of the system can be divided into two types: hardware and software components.

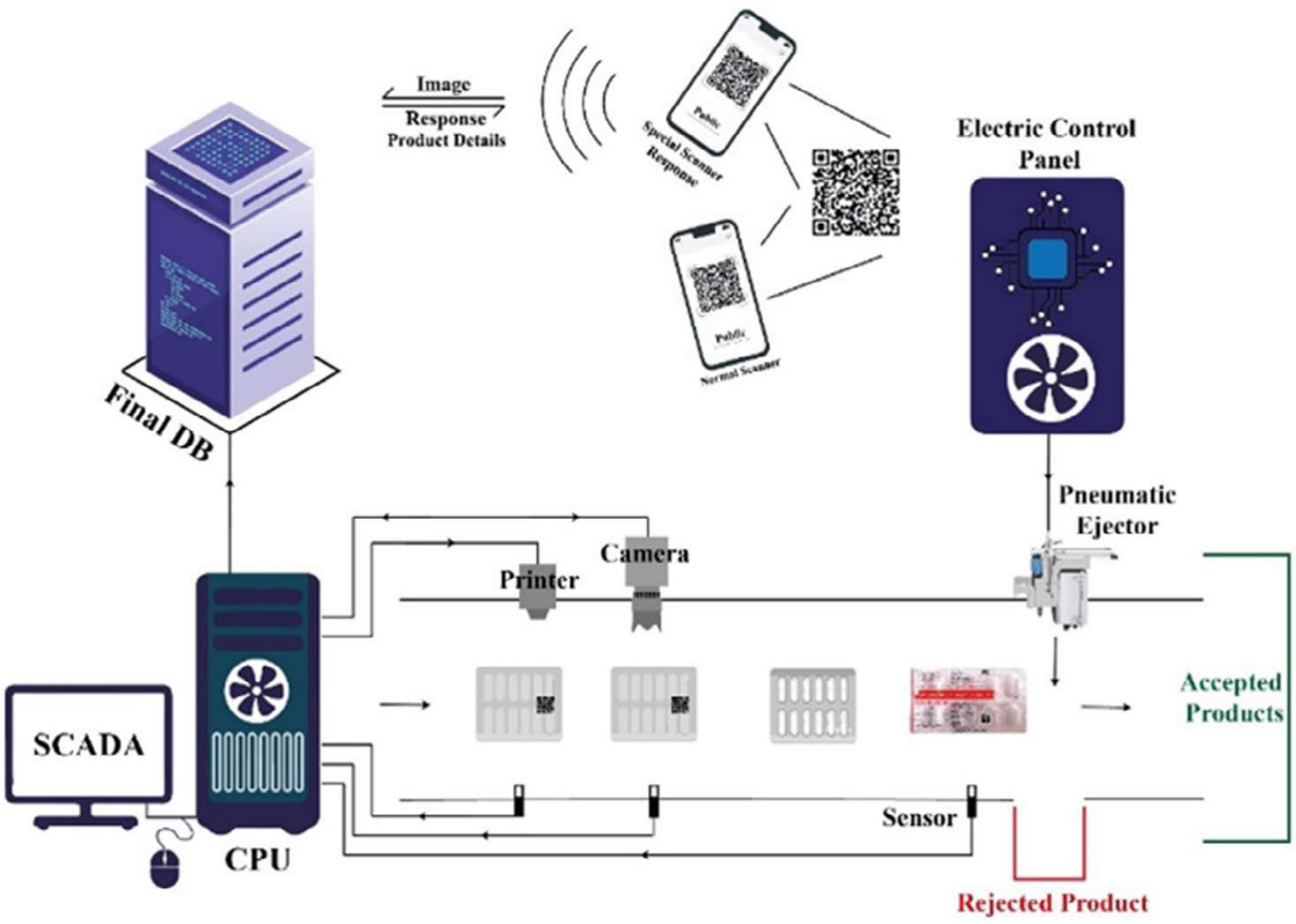

Fig. 3. Implementation architecture of the proposed system

### 3.1. Hardware components

The system was designed with consideration in mind that it can be installed with any in-line conveyor belt of pharmaceutical product manufacturing setup. The unique QR code is printed on the product packing using a thermal inject printer. A camera is used to capture the image of the printed QR codes for the verification of the printing. A proximity sensor is used to check the presence of the product to determine if the product needs to be accepted or rejected based on the image processing. The pneumatic ejector is used to eject the rejected product if there is any printing error. The printing and inspection system with the rejection unit is controlled by a Programmable Logic Controller (PLC) installed inside the electric control panel. The Supervisory Control and Data Acquisition (SCADA) PC is used to manage the production batch, generate, and inspect unique QR codes, and send the production data to the cloud server for the application layer. The SCADA PC is working as a connector for the line system and the main Database server. All the field devices, sensors, and electrical panels are connected to the machine SCADA PC. The cloud server is responsible for providing product data upon verified request from the end user and also for communicating with the machine setup. The hardware components are listed in Table 1.

Table 1: Hardware component used for the proposed system

| Component | Function | Specification |
| --- | --- | --- |
| Thermal inject printer | Printing the unique QR code | Print distance 0.5 - 2.0 mm (nozzle to print surface) |
| Camera | For image capture of the product | Cameras with Precision Time Protocol (PTP) |
| Pneumatic ejector | For separating the product with faulty QR codes | Capacity 30 to 600 GPM and discharge pressures up to 50 PSI |
| Proximity sensor | To identify the presence of the product | 3.3 to 5 Vdc Supply Input |
| SCADA PC | Central Processing Unit (CPU) | PowerEdge R350 |
| Final Database server | For storage of all data and working as a hub for information request and response | PowerEdge R750xs |

### 3.2. Software components

The SCADA PC uses customized software to generate and verify the protected QR code to send it to the cloud server. The cloud server runs verification software to authenticate the scanned QR code by the user using the customized Android application.

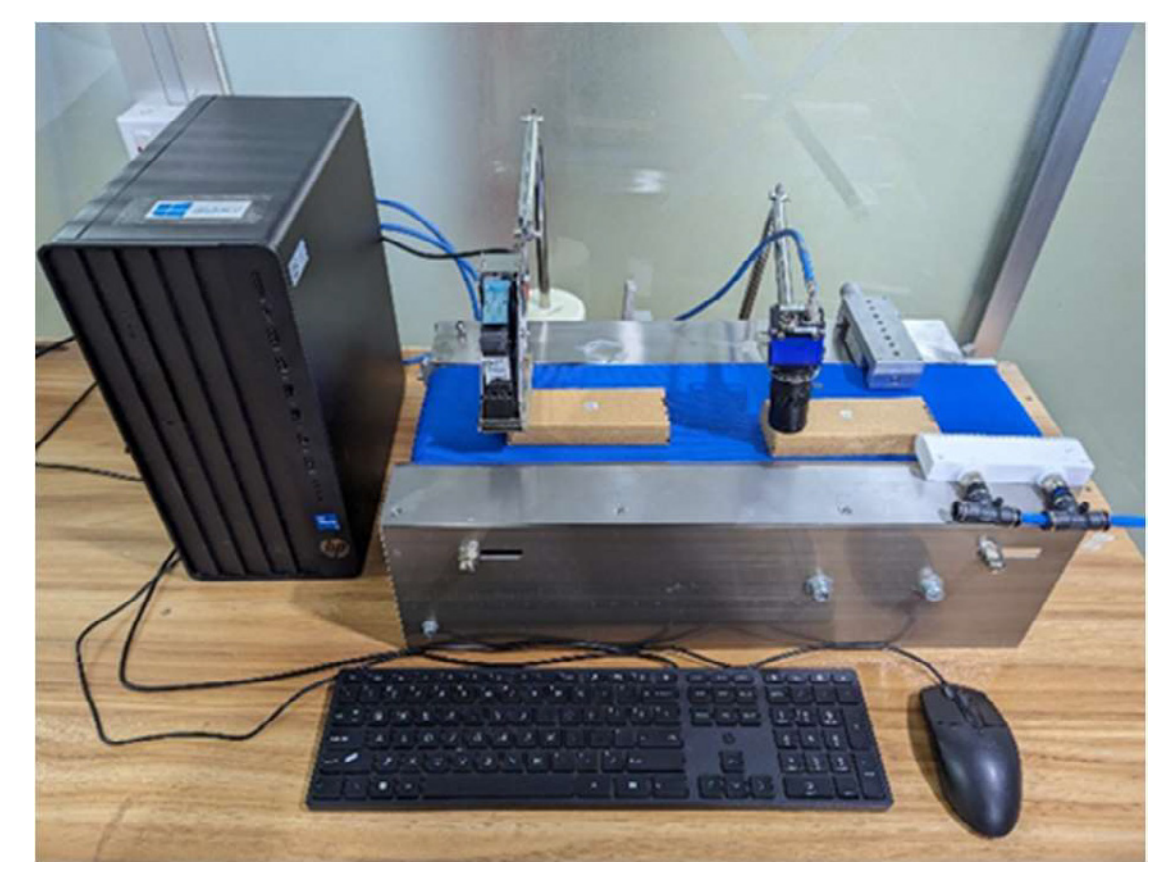

Fig. 4. Prototype of the proposed system for generating protected QR code

The prototype for the initial testing is shown in Fig. 4. For the initial testing the QR code was printed on sample medicine packages.

## 4. Results

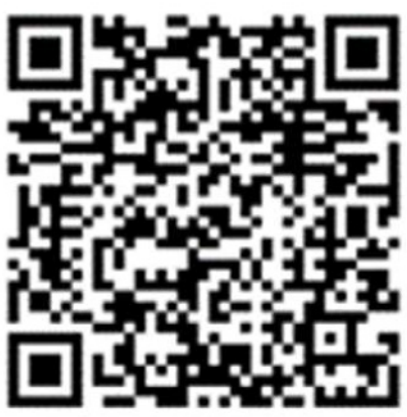

Fig. 5. Generated protected QR image encoded with unique and encrypted code

The sample output of the printed QR code is shown in Fig. 5. If the QR code is scanned by the conventional QR code scanner it will only show the public information: "Hello World!" but there is a private string: "KauHRusHAsm" inside the QR code that can only be decoded by the proposed system which ensures the protection of secret information. The encryption protects the private string even if the null separator is broken. The hidden modification process of encrypted private string adds a second layer of security to the data from reverse engineering. The private string is a primary key in the final database which is used to fetch all the information about the product stored in the database. The way the private string is encoded ensures that counterfeiters cannot decode the unique information to create counterfeit products. In the situation where the QR codes are duplicated by copying the image, the mobile application which is under development will track the location of the QR code scanning, and if there is any suspicious activity like the same QR codes are scanned multiple times or scanned in different locations the server system will notify the user about potential counterfeiting. Again, the production batch information will be collected during the printing process. Therefore, the system will know in which location the product is sold so if the scanning of the product is done in an unknown location user will also get a notification of the potential counterfeiting.

## 5. Conclusion

In this paper, a more systematic approach has been taken to prevent product counterfeiting by adopting a protected QR code and server mechanism in the pharmaceutical industry. The conventional and recent methods to eradicate this problem have also been discussed and yet the problem of counterfeiting is still increasing day by day. By implementing QR codes in each of the products and maintaining its traceability and uniformity through the server-based system, there will be less opportunity to counterfeit the product. The proposed system can be integrated into any pharmaceutical manufacturing environment with a few modifications. It is believed that by following the process mentioned in this paper, the counterfeiting issue can be lessened. However, more research needs to be performed in this sector and perfection can only be achieved by continued research work in this system. In future work, the proposed system will be implemented on a large scale to track and solve any feasibility issues. There will be always new challenges in the field of pharmaceutical manufacturing to prevent product counterfeiting. Hence, it is required to present a robust and sustainable anti-counterfeit solution to fight counterfeiting which will contribute towards sustainable manufacturing industry.